Expectancy and musical emotion

Effects of pitch and timing expectancy on musical emotion


Sauvé, S. A.[1], Sayed, A.[1], Dean, R. T.[2], Pearce, M. T.[1]

[1]Queen Mary, University of London

[2]Western Sydney University



Author Note

Correspondence can be addressed to Sarah Sauvé at s.a.sauve@qmul.ac.uk

School of Electronic Engineering and Computer Science

Queen Mary University of London, Mile End Road

London  E1 4NS

United Kingdom

+447733661107


Biographies

S Sauve: Originally a pianist, Sarah is now a PhD candidate in the Electronic Engineering and Computer Science department at Queen Mary University of London studying expectancy and stream segregation, supported by a college studentship.



A Sayed: Aminah completed her MSc in Computer Science at Queen Mary University of London, specializing in multimedia.

R.T. Dean: Roger is a composer/improviser and researcher at the MARCS Institute for Brain, Behaviour and Development.  His research focuses on music cognition and music computation, both analytic and generative.

M.T. Pearce: Marcus is Senior Lecturer at Queen Mary University of London, director of the Music Cognition and EEG Labs and co-director of the Centre for Mind in Society.  His research interests cover computational, psychological and neuroscientific aspects of music cognition, with a particular focus on dynamic, predictive processing of melodic, rhythmic and harmonic structure, and its impact on emotional and aesthetic experience. He is the author of the IDyOM model of auditory expectation based on statistical learning and probabilistic prediction.



Abstract

Pitch and timing information work hand in hand to create a coherent piece of music; but what happens when this information goes against the norm? Relationships between musical expectancy and emotional responses were investigated in a study conducted with 40 participants: 20 musicians and 20 non-musicians. Participants took part in one of two behavioural paradigms measuring continuous expectancy or emotional responses (arousal and valence) while listening to folk melodies that exhibited either high or low pitch predictability and high or low onset predictability. The causal influence of pitch predictability was investigated in an additional condition where pitch was artificially manipulated and a comparison conducted between original and manipulated forms; the dynamic correlative influence of pitch and timing information and its perception on emotional change during listening was evaluated using cross-sectional time series analysis. The results indicate that pitch and onset predictability are consistent predictors of perceived expectancy and emotional response, with onset carrying more weight than pitch. In addition, musicians and non-musicians do not differ in their responses, possibly due to shared cultural background and knowledge. The results demonstrate in a controlled lab-based setting a precise, quantitative relationship between the predictability of musical structure, expectation and emotional response.

Effects of pitch and timing expectancy on musical emotion

Music is capable of inducing powerful physiological and psychological emotional states (Bittman et al., 2013; Castillo-Pérez, Gómez-Pérez, Velasco, Pérez-Campos, & Mayoral, 2010). For example, the practice of music therapy stemmed from the observation that music can have a positive emotional impact (Khalfa, Bella, Roy, Peretz, & Lupien, 2003; Pelletier, 2004). However, many studies of emotion induction by music have simply investigated which emotions are induced rather the psychological mechanisms that account for why these emotions occur (Juslin & Västfjäll, 2008). The present research aims to address this omission by examining one theorized psychological mechanism of musical emotion induction in isolation. Although factors such as personality, age and gender have an influence (Rentfrow & Gosling, 2003), we focus here on the properties of music that are involved in emotion induction.

While there is general consensus that music can elicit emotional responses (see Juslin & Sloboda, 2001, for an extensive review), why and how it does so is less clear. Juslin & Vastfjall (2008) describe six potential psychological mechanisms that might explain how emotions are induced through music: (1) brain stem reflexes, (2) evaluative conditioning, (3) emotional contagion, (4) visual imagery, (5) episodic memory, and (6) musical expectancy. Hearing a sudden loud or dissonant event causes a change in arousal (brain stem reflex) whereas a piece repetitively paired with a positive, or negative, situation will create a positive, or negative, emotional reaction (evaluative conditioning).  Emotional contagion is the induction of emotion through mimicry of behavioural or vocal expression of emotion; for example shorter durations and ascending pitch contours tend to reflect happiness while longer durations and descending pitch contours communicate sadness.  Visual imagery refers to the mental imagery evoked by the music, which can have positive or negative affect.  Finally, pairing between a sound and a past



event can trigger the emotion related to that event when the sound is subsequently heard (episodic memory). The present study focuses on musical expectancy, while controlling for all other potential mechanisms proposed above. Meyer (1956) argues that emotion is generated through musical listening because listeners actively generate predictions reflecting what they expect to hear next (see also Huron, 2006). Unexpected events are surprising and associated with an increase in tension while expected events are associated with resolution of tension (e.g. Gingras et al., 2016). According to this account, surprising events generally evoke high arousal and low valence (Egermann et al., 2013; Koelsch, Fritz, & Schlaug, 2008; Steinbeis, Koelsch, & Sloboda, 2006). However, listeners familiar with a piece of music can come to appreciate an event that has low expectancy through an appraisal mechanism, resulting in a high valence response (Huron, 2006). This apparent contradiction highlights the importance of isolating the psychological mechanisms behind musical emotional induction. There are also different influences on musical expectation (Huron, 2006). Schematic influences reflect general stylistic patterns acquired through extensive musical listening to many piece of music while veridical influences reflect specific knowledge of a familiar piece of music. Dynamic influences reflect dynamic learning of structure within an unfamiliar piece of music (e.g. recognizing a repeated motif). Listening to new, unfamiliar music in a familiar style engages schematic and dynamic mechanisms, the former reflecting long-term learning over years of musical exposure and the latter short-term learning within an individual piece of music. Both these long- and short-term mechanisms can be simulated as a process of statistical learning and probabilistic generation of expectations (Pearce, 2005). Furthermore, these may be different for musicians and non-musicians due to extensive exposure and training in a particular style (Juslin & Vastfjall, 2008).



We now consider the properties of musical events for which expectations are generated.

Prominent among such properties are the pitch and timing of notes and we consider each in turn.

Music theorists have described musical styles as structurally organised, reflecting well

formalised rules that constitute a kind of grammatical syntax (Lerdahl & Jackendoff, 1983). In

the tradition of Western tonal music, compositions usually follow these rules by adhering to a

tonal structure and enculturated listeners are able to identify when a piece of music infringes

tonal rules based on their exposure in everyday music listening (Carlsen, 1981; Krumhansl,

Louhivuori, Toiviainen, Järvinen, & Eerola, 1999; Trainor & Trehub, 1992).  Two kinds of

models have been developed to explain listeners' pitch expectations: first, models that include

static rules; and second, models that focus on learning.  An influential example of a rule-based

model is the Implication-Realization (IR) model, developed by Eugene Narmour (1991) which

includes rules defining the expectedness of the final note in a sequence of three notes, where the

first pair of notes forms the implicative interval and the second pair of notes the realized interval.

The size and direction of the implicative interval sets up expectations of varying strengths for the

realized interval. While the original IR model contained five bottom-up rules of melodic

implication, Schellenberg (1997) reduced the five bottom-up rules of the IR model to two: *pitch

proximity* and *pitch reversal*. For example, according to the rules of *pitch reversal*, a small

interval implies another small interval in the same direction while a large interval implies a

subsequent small interval in the opposite direction. Such patterns reflect actual patterns in

existing Western music (Huron, 2006; Thompson & Stainton, 1996) suggesting the possibility

that listeners might learn these patterns through experience.

Statistical learning is a powerful tool for explaining the acquisition of pitch expectations

in music, where common sequential patterns are learned through incidental exposure (Huron,



2006; Pearce, 2005; Saffran, Johnson, Aslin, & Newport, 1999) making them more predictable to the exposed listener.  For example, a perfect cadence is found at the end of the vast majority of Western classical music, where movement from dominant to tonic is the strongest form of closure possible in this style. Through repeated exposure to this pattern, a dominant penultimate chord strongly implies a tonic chord for an enculturated listener (Huron, 2006).  IDyOM (Pearce, 2005) is a computational model of auditory expectation that harnesses the power of statistical learning.  It learns the frequencies of variable-order musical patterns from a large corpus of music (via the long-term model, or LTM) and from the current piece of music being processed (via the short-term model, or STM) in an unsupervised manner and generates probabilistic predictions about the properties of the next note in a melody given the preceding melodic context. The information content (negative log probability; IC) of an event, given the model, reflects the unexpectedness of the event in context. Low information content corresponds to high expectedness while high information content corresponds to low expectedness.

Temporal regularities are also learned through exposure (Cirelli, Spinelli, Nozaradan, & Trainor, 2016; Hannon & Trehub, 2005; Hannon & Trehub, 2005a; Hannon, Soley, & Ullal, 2012). Western music is dominated by beat patterns in divisions of two, and to a lesser extent, divisions of three, and compositional devices much as syncopation and hemiola (three beats in the time of two) are used to manipulate listener's temporal expectations. The dynamic attending theory (Jones & Boltz, 1989) posits that entrainment to a beat results in attentional focus being directed at time intervals where a beat is expected to occur, such that longer entrainment times result in stronger predictions and more focused attention. This was supported using a pitch discrimination task where participants performed better on pitch discrimination when target pitches fell on expected time points, as a result of entrainment to a series of isochronous



distractor tones (Jones, Moynihan, MacKenzie, & Puente, 2002; though see Bauer, Jaeger, Thorne, Bendixen, & Debener, 2015 for conflicting evidence). We propose that temporal rules can also be explained by statistical learning, as implemented in IDyOM (Pearce, 2005). In the same way as pitch, and various other musical surface structures, onset and inter-onset interval (IOI) can be predicted by IDyOM as it learns from a given corpus and a given piece. This is equivalent to estimating a distribution over possible future onset times, given the preceding sequences of events. Since pitch and temporal structures generate distinct expectancies, we explore the influence of each as a potential emotional inducer using both correlational and causal methods (while allowing for the possibility of interactions between pitch and timing).

Musical expectancy as a mechanism for the induction of emotion in listeners has been studied in an ecological setting: Egermann et al. (2013) asked 50 participants to attend a live concert, during which 6 flute pieces were played. These pieces spanned various musical styles and levels of pitch expectancy. Three areas of measure were taken: subjective responses (i.e. the arousal levels or the ratings of musical expectancy which changed continuously throughout the piece), expressive responses (using video and facial EMG) and peripheral arousal measured by skin conductance, heart rate, respiration and blood volume pulse. IDyOM (Pearce, 2005) was used to analyse pitch patterns of the music and predict where listeners would experience low expectancy. Results suggested that expectancy had a modest influence on emotional responses, where high IC segments led to higher arousal and lower valence ratings as well as increases in skin conductance and decreases in heart rate as compared to low IC segments while no event-related changes were found in respiration rate or facial electromyography (EMG) measures; however, this study was conducted in an ecologically valid, thus non-controlled environment where participants could have focused on something other than the music. For example, visual



aspects of performance are highly important to emotional engagement in live music settings

(Thompson, Graham, & Russo, 2005; Vines, Krumhansl, Wanderley, & Levitin, 2006).

Furthermore, other potential emotion inducing mechanisms, as proposed by Juslin & Vastfjall

(2008) were not explicitly controlled for and effects of temporal expectancy on emotional

responses were not considered.

The current study is designed to investigate pitch and temporal musical expectancy in a

restricted environment which controlled for many other potential emotional mechanisms (Juslin

& Vastfjall, 2008).  Brain stem reflexes are controlled for by maintaining equal tempo, intensity

and timbre across all musical excerpts.  Evaluative conditioning and episodic memory are

controlled for by presenting unfamiliar musical excerpts, so that expectation ratings and

emotional reactions are not confounded by previous experience with the music.  Potential effects

of emotional contagion are controlled for in the analysis by including pitch and IOI as predictors

of subjective ratings as well as pitch and IOI predictability (i.e. higher mean pitch and shorter

IOI could result in higher arousal and valence ratings, regardless of expectancy). Finally,

irrelevant visual imagery cannot be categorically avoided but the rating tasks are expected to

require enough cognitive load to render it unlikely. Furthermore, to the extent that visual imagery

is variable between individuals, averaging across participants should remove its influence.

This research aims to address three questions.  First, do the predictability of pitch and

timing (as simulated by IDyOM) have an effect on listeners' expectations and emotional state,

and can we causally influence this effect with explicit manipulation of the stimuli?  We

hypothesize that the degree of musical expectancy for pitch (based on pitch interval) and

temporal (based on inter onset interval) structures, as predicted objectively by information

content provided by IDyOM, will have an effect on emotion as measured by the arousal-valence



model (Russell, 2003), where unexpected events will invoke negative valence and cause an increase in arousal and expected events will invoke positive valence and decreased arousal. We also hypothesize that when both pitch and timing are either expected or unexpected, the emotional response will be more extreme than in conditions of mixed expectedness. Furthermore, direct manipulation of pitch expectancy while keeping temporal expectancy and all other musical features constant is expected to produce the predicted changes in ratings (i.e. transforming unexpected pitches to expected pitches will decrease unexpectedness and arousal, and increase valence ratings).

Second, how do pitch and timing predictability combine to influence expectation and emotion? Though the combination of pitch and timing in music perception has been a research interest for decades (Boltz, 1999; Duane, 2013; Jones, Boltz, & Kidd, 1982; Palmer & Krumhansl, 1987; Prince, Thompson, & Schmuckler, 2009), no clear conclusions can be drawn as findings regarding this question have low agreement and seem highly dependent on the choice of stimuli, participants and paradigm. For example, while Prince et al. (2009) suggest that pitch is more salient, results from Duane's (2013) corpus analysis suggest that timing is the most reliable predictor of streaming. While this study uses monophonic melodies, it could be argued that if salience is linked to complexity (Prince et al., 2009), then for melodies where pitch or timing are highly predictable (low complexity), the predictable feature will be more salient than its unpredictable counterpart. For melodies where pitch and timing are relatively equally predictable or unpredictable, their relative importance currently remains unknown.

Finally, is there a difference in the effect of pitch and timing predictability on expectation and emotional responses between musicians and non-musicians? The effect of musical training will be evaluated by comparing the responses of musicians and non-musicians, with the



expectation that musicians will have higher expectation ratings and more extreme emotional responses to pitch and timing violations due to training (Strait, Kraus, Skoe & Ashley, 2009), where Western musical patterns are more familiar, resulting in violations of these patterns eliciting stronger responses.

## Method

### Participants

40 participants (22 female, 18 male; age range 14-54 were recruited from universities, secondary school and colleges for this experiment: 20 were musicians (mean 3.6 years of musical training, range 1 – 12 years) and 20 were non-musicians (0 years of musical training).

### Stimuli

The stimuli consisted of 32 pieces of music in MIDI format rendered to audio using a piano timbre: 16 original melodies and 16 artificially-manipulated melodies. Original melodies were divided into the following four categories of predictability: predictable pitch and predictable onset (PP), predictable pitch and unpredictable onset (PU), unpredictable pitch and predictable onset (UP) and unpredictable pitch and unpredictable onset (UU). The artificial melodies were created by changing the pitch predictability of each melody so that PP became aUP, UU became aPU, PU became aUU and UP became aPP, where *a* denotes artificial. All melodies were presented at the same intensity, which was held constant for the duration of all melodies.

**Original melodies.** The sixteen original melodies were selected from a group of nine datasets, totalling 1834 melodies (see Table 1 for details), all from Western musical cultures to avoid potential cultural influences on expectancy ratings (Hannon & Trehub, 2005; Palmer & Krumhansl, 1990). All nine datasets were analysed by IDyOM for target viewpoints pitch and



onset with source viewpoints pitch interval and scale degree (linked), and inter-onset-interval (IOI) respectively. Both STM and LTM models were engaged; the LTM model was trained on three datasets of Western music, described in Table 2. There was no resampling within the test datasets.

The 1834 melodies were divided into four categories based on high or low pitch or onset information content (IC). Melodies were considered predictable if they had a lower IC than the mean IC of all samples and unpredictable if the IC was greater than the mean IC of all samples. Four melodies from each category were selected as the most or least predictable by finding maximum and minimum IC values as appropriate for the category; these are the original sixteen melodies. Melodies in the PP, PU, UP and UU categories had mean pitch IC values ranging from 1.37-1.85, 2.22-2.43, 2.83-5.24 and 2.61-2.78 respectively, mean onset IC values ranging from .80-.92, 2.49-4.34, 1.13-1.32 and 4.20-4.39 respectively, mean raw pitch values ranging from 66.85-70.17, 66.05-70.23, 68.67-72.76 and 64.40-71.63 respectively and mean IOI values ranging from 12.71-21.28, 21.41-69.81, 13.84-21.69 and 21.53-64.00 respectively, where a quarter note equals 24. Notice that categories with unpredictable onset have higher average IOI values; this potential confound is discussed below (see Appendix A).

[Table 1]

[Table 2]

**Artificial melodies.**

The sixteen artificial melodies were created as follows. For each original melody, the notes with the highest (for PP and PU) or lowest (for UP and UU) information content were selected for replacement. The notes were replaced with another note from the same melody which shares the same preceding note as the original note in that melody. If several instances of



such a note pair existed, the associated IC values were averaged. If several such replacement notes existed, the one with the lowest (for UP and UU) or highest (for PP and PU) information content was selected to replace the original note. Where no such replacement existed, the key of the melody was estimated using the Krumhansl-Schmuckler key-finding algorithm (Krumhansl, 1990) using key profiles updated by Temperley (1999) and the replacement was selected as the scale degree with highest (for UP and UU) or lowest (for PP and PU) tonal stability. All notes labelled as having extremely high or low IC were replaced by a pitch with a less extreme IC.

Melodies in the aPP, aPU, aUP and aUU categories had mean pitch IC values ranging from 3.49-5.50, 4.20-4.56, 4.13-6.59 and 2.79-3.80 respectively and mean raw pitch values ranging from 64.88-69.80, 67.05-73.18, 64.05-67.76 and 66.78-72.89 respectively. Mean onset IC and mean raw IOI values were unchanged from the corresponding original stimulus predictability category (e.g. aPP has the same mean IOI IC and IOI values as UP). Figure 1 illustrates the mean information content of all 32 melodies.

[Figure 1]

### Procedure

Participants were semi-randomly allocated to one of four (between-subjects) conditions: they were either a musician or a non-musician and, within these groups, randomly assigned to rate either expectancy or emotion (arousal and valence). The experiment was run on a Samsung Galaxy Ace (Android 2.3.6). The information sheet was presented and informed consent gathered; detailed step-by-step instructions were then presented to participants. Regardless of condition, there was a mandatory practice session: participants heard two melodies and answered the questions appropriate to the condition they were assigned to (either expectancy rating or arousal and valence rating). Once the practice session was completed, the experimental app was



loaded. Participants entered a unique ID number provided by the experimenter and responded to

a short musical background questionnaire. Participants then heard the 32 musical excerpts (mean

duration 18.34 s) presented in random order without pause or repeat and performed the

appropriate ratings by continuously moving a finger on the screen. Those in the expectancy

rating condition reported expectancy on a 7-point integer Likert scale, where 1 was *very expected*

and 7 was *very unexpected*. Those in the arousal/valence condition rated induced arousal

(vertical) and valence (horizontal) on a two-dimensional arousal/valence illustration (Russell,

2003). Responses, in integers, were collected at a 5Hz sample rate (200ms) (Khalfa et al., 2002).

The rating systems used were: Expectancy: $1 - 7$ (expected - unexpected); Arousal: $0 - 230$

(calm - stimulating); Valence: 0 - 230 (unpleasant – pleasant).

**Statistical analysis**

For each type of rating, a melody-level analysis was performed. For each melody, a mean

expectancy rating was calculated at every time point across the musician and non-musician

groups (10 responses per group). Linear multiple regression modelling was used to evaluate the

impact of *time* (point in time at sampling rate of 200ms), *musical training* (musician or non-

musician), *stimulus modification* (original or artificial), *stimulus predictability*

(predictable/unpredictable pitch/onset), *pitch* and *IOI* on mean expectancy ratings by comparing

a model with each predictor to a model containing only an intercept. Pitch and IOI are included

as predictors to control for potential confounding effects of musical contagion. While musical

training and stimulus modification were simple factors, contrasts were set for stimulus

predictability to compare all pairs of categories to each other. Interactions were not considered

due to the difficulty of interpretation in such a complex model.



Aside from fundamental analytical summary statistics which deal directly with the causal interventions, we employed cross-sectional time series analysis (CSTSA) similarly to Dean et al. (Dean, Bailes, & Dunsmuir, 2014a) to evaluate the predictive impact effects of *pitch IC*, *onset IC*, *stimulus predictability* (predictable/unpredictable), *stimulus modification* (none/artificial), *musical training* and individual differences modelled by random effects on participants' ratings of expectedness, arousal and valence. CSTSA takes account of the autoregressive characteristic of music and the continuous responses of the participants. Pitch IC and onset IC predictors were both scaled to values between 0 and 1 to allow for direct comparison of model coefficients in analysis. A predictor of combined *pitch and onset IC* was also tested, replacing the individual *pitch IC* and *onset IC* predictors. In practice, CSTSA is a mixed-effects model, fitted with maximum random effects as per Barr et al. (Barr, Levy, Scheepers, & Tily, 2013) and fixed effects to account for autocorrelation (lags of endogenous variables, i.e. ratings, denoted by P), and exogenous influence (i.e. information content and its lags, denoted by L). Only optimal models are presented below, selected based on BIC, confidence intervals on fixed effect predictors, log likelihood ratio tests between pairs of models, correlation tests between models and the data, and the proportion of data squares fit.

## Results

### Melody level analysis

In this section we describe analyses of the mean ratings melody by melody and participant by participant: these are discontinuous data, and the experiment manipulated the pitch expectancy of the original melodies, to provide a causal test of its influence. Mean ratings are shown in Figure 2 and important comparisons are highlighted in Figure 3.



**Expectancy ratings.** There were significant effects of time ($F_{(1)} = 742.01$, $p < .0001$), musical training where musicians rated melodies with higher unexpectedness (musicians mean = 4.40; non-musicians mean = 4.16; $F_{(1)} = 73.12$, $p < .0001$), stimulus modification where modified melodies were rated as more unexpected (original melodies mean = 3.92; modified melodies mean = 4.65; $F_{(1)} = 569.75$, $p < .0001$), and stimulus predictability where more predictable melodies were rated with lower unexpectedness than unpredictable melodies (PP melodies mean = 3.48; PU melodies mean = 4.71; UP melodies mean = 3.92; UU melodies mean = 4.66; $F_{(3)} = 251.58$, $p < .0001$) on mean expectancy ratings. We also investigated the effect of stimulus predictability on ratings for original and modified melodies separately, where means for PP, PU, UP and UU melodies were 1.88, 4.47, 3.58 and 5.19 respectively ($F_{(3)} = 1866.2$, $p < .0001$) and for aPP, aPU, aUP and aUU melodies were 4.27, 4.16, 5.29 and 4.96, respectively ($F_{(3)} = 264.36$, $p < .0001$). Pitch was not a significant predictor, $F_{(1)} = 1.37$, $p = .24$, while IOI was, $F_{(1)} = 22.04$, $p < .0001$.

**Arousal ratings.** There were significant effects of time ($F_{(1)} = 127.32$, $p < .0001$), musical training where musicians rate melodies as more arousing overall as compared to non-musicians (musicians mean = 118.16; non-musicians mean = 112.90; $F_{(1)} = 25.30$, $p < .0001$), stimulus predictability where more predictable melodies were rated as more arousing (PP melodies mean = 151.73; PU melodies mean = 109.45; UP melodies mean = 128.86; UU melodies mean = 95.95; $F_{(3)} = 667.31$, $p < .0001$), pitch ($F_{(1)} = 62.95$, $p < .0001$) and IOI ($F_{(1)} = 6.59$, $p = .01$). There was no effect of stimulus modification (original melodies mean = 115.83; modified melodies mean = 115.27; $F_{(1)} = .62$, $p = .42$). Stimulus predictability was also a significant predictor when original and artificial melodies' ratings were investigated separately, with ratings for PP, PU, UP and UU melodies averaging 138.62, 111.14, 121.07 and 100.79



respectively, F (3) = 210.16, p < .0001, and aPP, aPU, aUP and aUU melodies averaging 137.10, 91.56 144.96 and 107.83, respectively, F (3) = 556.76, p < .0001.

**Valence ratings.** There were significant effects of time (F (1) = 6.29, p = .01), musical training where musicians overall rated melodies as having lower valence (musicians mean = 81.26; non-musicians mean = 84.08; F (1) = 5.38, p = .02), stimulus modification where original melodies had more positive valence than artificial melodies (original melodies mean = 91.20; artificial melodies mean = 74.33; F (1) = 206.84, p < .0001), stimulus predictability where more predictable melodies are rated more positively than unpredictable melodies (PP melodies mean = 109.87; PU melodies mean = 74.00; UP melodies mean = 87.00; UU melodies mean = 70.02; F (3) = 224.81, p < .0001), pitch (F (1) = 44.79, p < .0001), and IOI (F (1) = 210.44, p < .0001). Stimulus predictability is also a significant predictor when investigating original and artificial melodies separately, where PP, PU, UP and UU melodies have mean arousal ratings of 171.90, 77.96, 94.59 and 44.46 respectively, F (3), 1582.6, p < .0001 and aPP, aPU, aUP and aUU melodies have mean ratings of 78.98, 93.21, 45.66 and 70.19 respectively, F (3) = 276.84, p < .0001.

[Figure 2]

[Figure 3]

**Cross-sectional time series analysis**

Here we present the analyses of the continuous time series data resulting from participants' ongoing responses during listening to the melodies.

Expectancy, arousal and valence ratings were modelled separately using mixed-effects autoregressive models with random intercepts on *participant ID* and *melody ID* as well as random slopes on the fixed effect predictor with the largest coefficient before slopes were added.



Fixed effects predictors were *time*, *musical training*, *stimulus predictability*, *stimulus modification*, autoregressive lags of up to 15 (equivalent of 3 seconds) and exogenous lags of pitch and onset information content of up to 15.  A combined pitch and onset information predictor was also tested to evaluate whether a combined measure superseded the separate pitch and onset information content predictors.  Maximum lags were selected based on previously reported rate of change of emotional responses (Juslin & Västfjäll, 2008) as well as precedent in this type of analysis (Dean et al., 2014a).  Pitch and IOI were subsequently added as fixed-effect predictors to investigate the potential confounding effects of musical structure affecting ratings through an emotional contagion mechanism. See figures 4 and 5 for an illustration of the variance fitted by random effects, and the fit of the models for a selection of melodies and participants.

**Expectancy ratings.** The best CSTSA model for expectancy ratings is summarized in Table B1 in Appendix B. In this model, while autoregression and random effects were duly considered, an effect of musicianship was still clearly observed in addition to pitch IC and onset IC and a selection of their lags. Thus the model included random intercepts and random slopes for L1pitchIC on melody ID and participant ID as well as fixed effects of musicianship, L = 0-1, 7-8 of pitch IC, L = 0-2, 10, 12 of onset IC and P = 1-2, 4-6, 15 of autoregression.  All predictors were significant, as Wald 95% confidence intervals did not include zero.  The addition of stimulus predictability as a fixed effect did not improve the model, $\chi^2$ (3) = 1.80, p = .61 while musicianship and stimulus modification did, $\chi^2$ (2) = 13.36, p = .001 and $\chi^2$ (1) = 3.91, p = .04 respectively.  The further addition of pitch and IOI significantly improved the model, $\chi^2$ (2) = 409.33, p < .0001, and removed stimulus modification as a significant predictor.  Combined pitch



and onset information content with lags of pitch and onset from the best model outlined above was significantly worse, $\chi^2$ (6) = 972.6, p < .0001.

A correlation test between the data and the model is highly significant, with correlation .93, t (82486) = 783.09, p < .0001. A proportion of data squares fit test is also high, with the model explaining 98% of the data. While this particular model did not converge, a model without random slopes removed did converge where all fixed effects were significant, model fit was equally good (correlation test: .93, t (82486) = 780.53, p < .0001; proportion of data squares fit: 98%) and the inclusion of slopes improved the model significantly; therefore random slopes were reinserted into the best model as per the experimental design (Barr et al., 2013).

**Arousal ratings.** The best CSTSA model for arousal ratings is summarized in Table B2 in Appendix B. This model revealed stimulus predictability as a significant predictor of arousal ratings in addition to pitch IC and onset IC and a selection of their lags when autoregression and random effects were considered. The model included random intercepts and random slopes for L1onsetIC on melody ID and participant ID as well as fixed effects L = 0-1, 6-8, 10-13, 15 of pitch IC, L = 0-4, 7, 10, 12-15 of onset IC and P = 1, 3, 5-6, 15 of autoregression. All predictors were significant, as Wald 95% confidence intervals did not include zero. The addition of musicianship and stimulus modification as fixed effects did not improve the model, $\chi^2$ (2) = .60, p = .74 and $\chi^2$ (2) = 1.72, p = .42 respectively while stimulus predictability did, $\chi^2$ (2) = 14.91, p = .0005. The further addition of pitch and IOI significantly improved the model, $\chi^2$ (2) = 178.89, p < .0001, where both are significant predictors of arousal ratings. Combined pitch and onset information content with lags of pitch and onset from the best model outlined above was significantly worse, $\chi^2$ (13) = 4482.2, p < .0001.



A correlation test between the data and the model is highly significant, with correlation .96, t (80183) = 978.48, p < .0001.  A proportion of data squares fit test is also high, with our model explaining 98% of the data. While this particular model did not converge, a model without random slopes removed did converge where all fixed effects were significant, model fit was equally good (correlation test: .95, t (80183) = 959.73, p < .0001; proportion of data squares fit: 98%) and the inclusion of slopes improved the model significantly, $\chi^2$ (5) = 335.3, p < .0001; therefore random slopes were reinserted into the best model as per the experimental design (Barr et al., 2013).

**Valence ratings.** The best CSTSA model for valence ratings is summarized in Table B3 in Appendix B.  This model revealed significant effects of only pitch IC and onset IC and a selection of their lags when autoregression and random effects were considered.  The model included random intercepts and random slopes for L1onsetIC on melody ID and participant ID as well as fixed effects L = 0-1, 5, 8-9, 11-13, 15 of pitch IC, L = 0-1, 3-4, 10, 13 of onset IC and P = 0, 3-7, 9, 15 of autoregression.  All predictors were significant, as Wald 95% confidence intervals did not include zero.  The addition of musicianship, stimulus predictability and modification as fixed effects did not improve the model, $\chi^2$ (1) = .29, p = .58, $\chi^2$ (3) = 4.77, p = .18 and $\chi^2$ (1) = 3.46, p = .06 respectively.  The further addition of pitch and IOI significantly improved the model, $\chi^2$ (1) = 600.99, p < .0001, where both are significant predictors of arousal ratings.  Combined pitch and onset information content with lags of pitch and onset from the best model outlined above was significantly worse, $\chi^2$ (10) = 194.72, p < .0001.

A correlation test between the data and the model is highly significant, with correlation .94, t (80183) = 827.83, p < .0001.  A proportion of data squares fit test is also high, with our model explaining 98% of the data.  While this particular model did not converge, a



model without random slopes removed did converge where all fixed effects were significant, model fit was equally good (correlation test: .94, t (80183) = 959.73, p < .0001; proportion of data squares fit: 95%) and the inclusion of slopes improved the model significantly, $\chi^2$ (4) = 805.25, p < .0001; therefore random slopes were reinserted into the best model as per the experimental design (Barr et al., 2013).

[Figure 4]

[Figure 5]

## Discussion

The results provide answers to all three of our research questions. First, we find evidence that predictability of both pitch and temporal musical structure have an effect on listeners' expectancies and emotional reactions, and that these can be manipulated. Second, we find that temporal expectancy influences perception more strongly than pitch expectancy. Finally, we find that individual differences generally supersede effects of musical training (Dean et al., 2014a) and inter-melody differences were more substantial than differences between melody predictability groups (PP, UP, PU and UU) or manipulation type, where differences between predictability groups could nevertheless be detected in the discontinuous, melody-level analysis.

Using IDyOM (Pearce, 2005) to calculate average pitch and onset information content, we classified folk songs into four categories based on overall expectedness, where average pitch expectancy and average onset expectancy could be high or low. We also manipulated pitch expectancy to transform expected pitches into unexpected ones, and vice versa. The four melody categories resulted in different subjective ratings of expectancy, arousal and valence, where high pitch and onset information content (PP) resulted in high unexpectedness ratings, higher arousal and lower valence, low pitch and onset information content (UU) resulted in low unexpectedness



ratings, lower arousal and higher valence, and mixed high and low pitch and onset information

content (PU and UP) lay somewhere in between, where only the predictable pitch and onset (PP)

and unpredictable pitch and predictable onset (UP) categories were not different from each other

in arousal ratings.  This confirms previous evidence that statistical learning and information

content may influence listener expectancies (Pearce, Ruiz, Kapasi, Wiggins, & Bhattacharya,

2010; Pearce & Wiggins, 2006) and arousal and valence ratings of music (Egermann et al.,

2013).  Cross-sectional time series analysis support these results with excellent models

explaining between 93-96% of expectancy, arousal and valence ratings, all including pitch and

onset information content, and lags of these of up to 3s (Egermann et al., 2013) as predictors. We

additionally find that explicit causal manipulation of pitch expectancy – the modification of

selected pitches from high to low or from low to high expectancy – results in a change in ratings

in the expected direction.  For example, melodies transformed into the UP category (filled

triangle in Figure 2) are rated with higher unexpectedness ratings and lower valence than their

original PP counterparts (hollow square in Figure 2), yet are also different from the original UP

category (hollow triangle in Figure 2) melodies.  This effect is more pronounced for

expectedness and valence ratings than for arousal ratings, which can be explained by the

intentionally inexpressive nature of the stimuli.  Therefore, the manipulation of pitch expectancy

adds causal evidence to previous research by demonstrating a direct link between expectancy

manipulation and expectancy, arousal and valence ratings.

CSTSA also allows us to assess the relative contribution of pitch and onset information

content to expectancy, arousal and valence ratings.  We find that onset information content

coefficients are almost always approximately 1.1 to 4.3 times larger than pitch information

content coefficients for exactly (i.e. L1pitchIC and L1onsetIC) or loosely (i.e. L5pitchIC and



L6onsetIC) matching lags.  Furthermore, the sum of onset IC lag coefficients is far greater than the sum of pitch IC lag coefficients for arousal and valence rating models, while the sum of pitch IC lag coefficients is greater than onset IC lag coefficients for the expectancy ratings model (though absolute values of individual onset IC coefficients are greater than the pitch IC coefficients).  Incidentally, every model includes pitch IC and onset IC lags of 0 and 1, with little overlap beyond this, suggesting that processing time scales for both pitch and onset expectancy are similar soon after a particular note event and diverge after this.  This variation in time scales could also explain why a combined pitch and onset IC predictor did not replace the separate pitch IC and onset IC predictors.

Though analysis of mean ratings yielded a main effect of musical training, the amount of variance explained by musical background was superseded by the amount of variance explained by random effects on participant ID for arousal and valence ratings, indicating that though groups can be formed, individual strategies are more important to explain these ratings.  Though a large body of literature supports the existence of certain differences between musicians and non-musicians (Brattico, Näätänen, & Tervaniemi, 2001; Carey et al., 2015; Fujioka, Trainor, Ross, Kakigi, & Pantev, 2004; Granot & Donchin, 2002), similar research by Dean et al. (Dean et al., 2014a; Dean, Bailes, & Dunsmuir, 2014b) has also found that though there were differences between groups, individual differences explain more variance than musical background when rating arousal and valence of electroacoustic and piano music.  However, musical background did hold important predictive power for expectancy ratings, where musicians gave slightly higher ratings overall, showing greater unexpectedness.  Though one might at first expect musicians to have lower expectancy ratings overall due to expertise with musical patterns, the alternative is possible when considering work by Hansen & Pearce (2014),



who present evidence that musicians make more specific predictions (i.e., predictions that are lower in entropy or uncertainty) than non-musicians when listening to music.  It is possible that due to these more specific predictions, any violations were perceived as more unexpected, as opposed to the less specific predictions of a non-musician, which would result in less surprise when violated.  That being said, it is worth noting that the overall difference in ratings between musicians and non-musicians is small, with musicians' ratings being only .2 points higher.

Similarly, we found that the differences between individual melodies, as modelled by random intercepts and slopes on Melody ID, outweigh categories of stimulus predictability and stimulus modification in all but two cases: expectancy ratings, where stimulus modification was a significant predictor, and arousal ratings, where stimulus predictability was a significant predictor, such that PP > UP > PU > UU in terms of arousal ratings.  The predictive power of stimulus modification in the context of expectancy ratings can be explained by the overall higher pitch IC in artificial melodies, as shown in Figure 2.  This is likely due to the fact that the modifications were made by an algorithm and are therefore not as smooth as human-composed changes might have been.  As the original melodies already had relatively low IC, it would be difficult to keep mean IC as low or lower with the change of even one note, as this change could also have an effect on the IC of all subsequent notes in a given melody.

As for the importance of predictability in predicting arousal ratings, which was in the opposite direction to what was predicted based on previous empirical (Egermann et al., 2013; Steinbeis et al., 2006) and theoretical (Meyer, 1956;Huron, 2006) research, this could be explained by the potentially confounding effect of duration on ratings.  Our analysis revealed that note duration did indeed have a significant effect on ratings, where melodies with longer durations, corresponding to low onset expectancy, were rated as more unexpected, less arousing



and less pleasant.  The pattern of mean arousal ratings by stimulus predictability, with PP and UP (high onset expectancy) rated as more arousing than PU and UU (low onset expectancy) matches this interpretation, which is further supported by previous research establishing a positive correlation between tempo and arousal (Carpentier & Potter, 2007; Husain, Thompson, & Schellenberg, 2002). The significant effect of pitch on ratings is more surprising; a pattern of higher average pitch for PP and UP categories corresponds to lower unexpectedness ratings, higher arousal ratings and higher valence ratings for these categories as compared to PU and UU categories.  However, coefficients for pitch and IOI are smaller than almost all other predictors in expectancy, arousal and valence models, suggesting that their overall influence is minimal compared to pitch and onset IC on subjective expectancy and emotion responses.

Also similarly to Dean et al. (2014a), the use of CSTSA allows us to evaluate evidence for the presence of a common perceptual mechanism across all pieces of music heard.  To do this, predictors encoding melodies by stimulus predictability and modification were added to the basic models, where a null effect of these additional predictors would indicate that the type of melody does not matter and the listeners' ratings depend only on pitch and onset IC in all melodies.  In the case of valence ratings, neither stimulus predictability nor stimulus modification were found to provide any additional predictive power to the model, while stimulus modification was a helpful predictor for expectancy ratings and stimulus predictability for arousal ratings.  However, explanations were proposed for these results and we maintain that our data provides some support for a common perceptual mechanism across all melodies.

**Relative salience**

The question of relative perceptual weighting between musical parameters such as pitch, timing, structure, and harmony in music cognition is important but challenging and lacks a unified



explanation (Dibben, 1999; Esber & Haselgrove, 2011; Prince et al., 2009; Uhlig, Fairhurst, & Keller, 2013).  Generally, pitch or melody is considered the most salient aspect of a piece of music. Prince et al. (2009), for example, argue that there are many more possible pitches than there are rhythmic durations or chords; therefore, pitch takes more attentional resources to process and is more salient.  On the other hand, in a corpus analysis of eighteenth- and nineteenth-century string quartets, Duane (2013) found that onset and offset synchrony were the most important predictors of streaming perception of these string quartets, with pitch explaining half the variance that onset and offset synchrony did, and harmonic overlap explaining an almost insignificant amount.  Our results indicate that onset information content is more salient than pitch information content, though here we evaluate the perception of emotion alongside the subjective experience of expectancy, as opposed to auditory streaming.  Interestingly, work in cue salience outside of music explores the effect of predictability and uncertainty on salience (Esber & Haselgrove, 2011), with one model predicting increased salience for cues with high predictability (Mackintosh, 1975) and another model predicting increased salience for cues with high uncertainty (Pearce & Hall, 1980). Though contradictory, these models have each accumulated significant evidence and have more recently led to the development of both hybrid (Pearce & Mackintosh, 2010) and new unified models of cue salience (Esber & Haselgrove, 2011).  We considered the possibility that high and low uncertainty and pitch and onset lag coefficients interacted so that melodies with high pitch predictability (expectancy) and low onset predictability (PU) led to larger pitch IC coefficients than onset  IC coefficients, and vice versa. This effect was not found in the data (see Appendix C), so we conclude that in this particular paradigm, onset is the more salient cue overall.

**A mechanism for emotional induction**



Returning to the identified lack of research into specific mechanisms for emotional induction by music (Juslin & Västfjäll, 2008; Meyer, 1956), the present research makes a single but significant step towards isolating individual mechanisms.  The study explicitly controlled for four of the six proposed mechanisms and manipulated one while considering another as a covariate.  Brain stem reflexes, evaluative conditioning, episodic memory and visual imagery are controlled for by presenting novel stimuli with equal tempo, intensity and timbre alongside a rating task.  Emotional contagion, information conveyed by musical structure itself, was addressed by including pitch and duration values into our CSTSA models of the expectancy, arousal and valence ratings.  Though these were significant predictors, they carried less weight than the lags of information content predictors.  We examined musical expectancy by selecting stimuli with either high or low pitch and onset expectancy and additionally explicitly manipulated pitch expectancy, finding evidence for a consistent effect of pitch and onset expectancy on ratings of arousal and valence by musicians and non-musicians.  We additionally find that onset is more salient than pitch and that musicians give higher unexpected ratings than non-musicians, but group differences are overridden by individual differences on emotion ratings.  Potential future work includes the use of stimuli at less extreme ends of the expectancy spectrum to validate these findings, manipulating onset IC in addition to pitch IC, allowing the evaluation of dependencies between the two (see Palmer & Krumhansl, 1987), exploring interactions of predictability and entropy on salience cues in emotion ratings and investigating other potential emotional induction mechanisms in a similarly controlled way, working towards an integrated model of musical emotion induction and perception.

Tables

Table 1

*Details of the datasets used in stimulus selection.*

| Dataset | Description | Number of melodies | Mean events/composition |
|---|---|---|---|
| 2 | Chorale soprano melodies harmonized by J.S. Bach | 100 | 46.93 |
| 3 | Alsatian folk songs from the Essen Folk Song Collection | 91 | 49.40 |
| 4 | Yugoslavian folk songs from the Essen Folk Song Collection | 119 | 22.61 |
| 5 | Swiss folk songs from the Essen Folk Song Collection | 93 | 49.31 |
| 6 | Austrian folk songs from the Essen Folk Song Collection | 104 | 51.01 |
| 10 | German folk songs from the Essen Folk Song Collection: ballad | 687 | 40.24 |
| 15 | German folk songs from the Essen Folk Song Collection: kinder | 213 | 39.40 |
| 18 | British folk song fragments used in the experiments of Schellenberg (1996) | 8 | 18.25 |
| 23 | Irish folk songs encoded by Daiman Sagrillo | 62 | 78.5 |



Table 2

*Details of the training set used to train IDyOM.*

| Dataset | Description | Number of melodies | Mean events/composition |
|---------|-------------|--------------------|-------------------------|
| 0 | Songs & ballads from Nova Scotia, Canada | 152 | 56.26 |
| 1 | Chorale melodies harmonized by J.S. Bach | 185 | 49.88 |
| 7 | German folk songs | 566 | 58.46 |



**Figures**

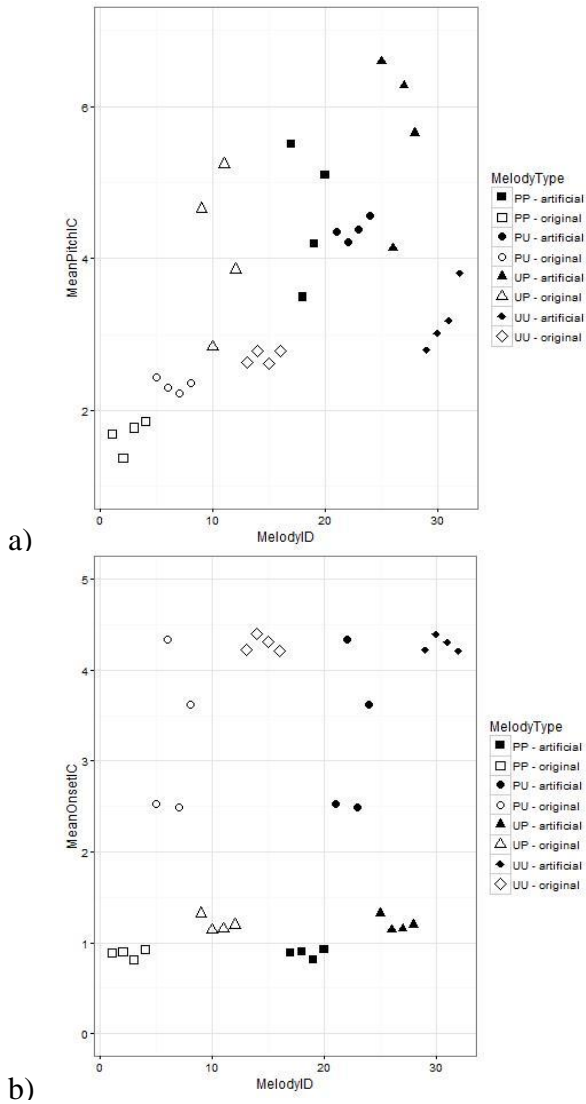

a)

b)

Figure 1. Mean (a) pitch IC and (b) onset IC of each melody plotted by stimulus predictability and modification, where original melodies are symbolized by empty symbols and artificial melodies by full symbols.



Musicians                                                    Non-musicians

Expectancy

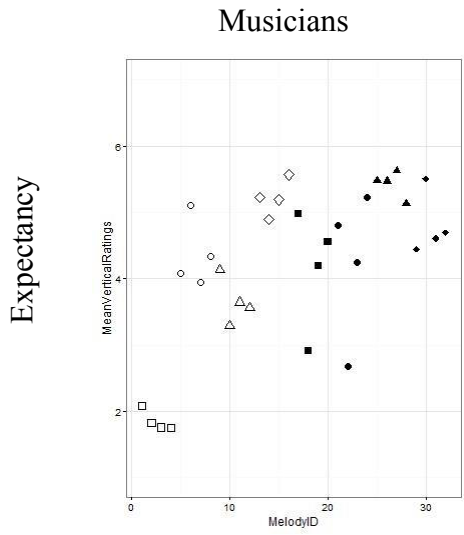 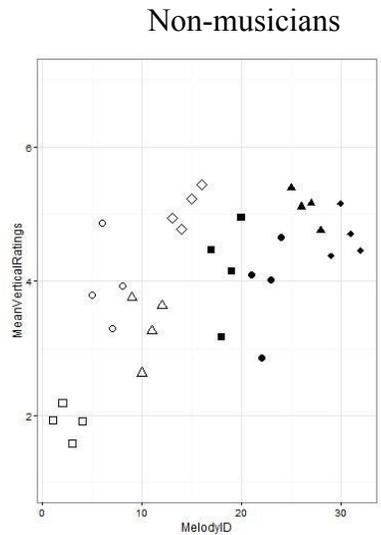

a)                                                          b)

Arousal

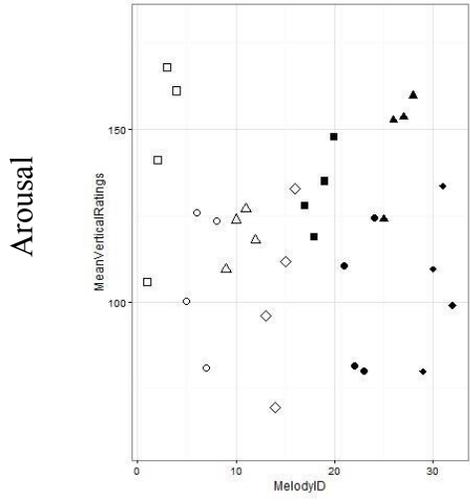 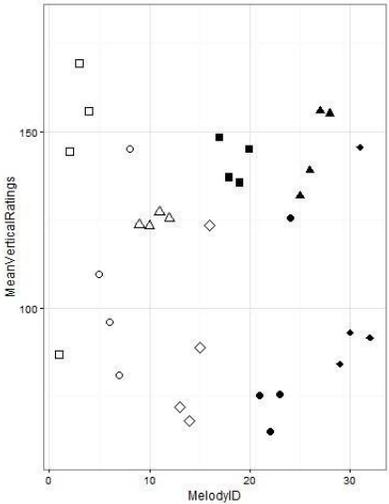

c)                                                          d)

Valence

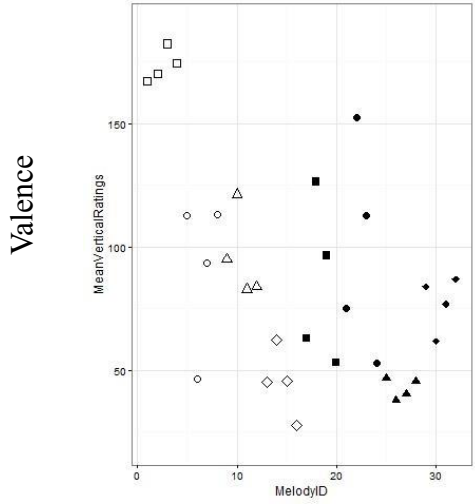 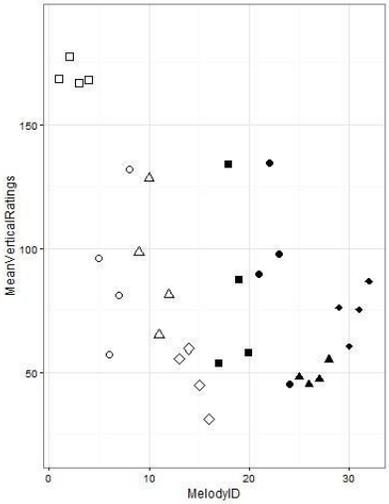

e)                                                          f)



Figure 2. Mean expectancy (a, b), arousal (c, d) and valence (e, f) ratings for each melody

for musicians (a, c, e) and non-musicians (b, d, f). Hollow shapes illustrate original melodies

while filled shapes illustrate artificial melodies.



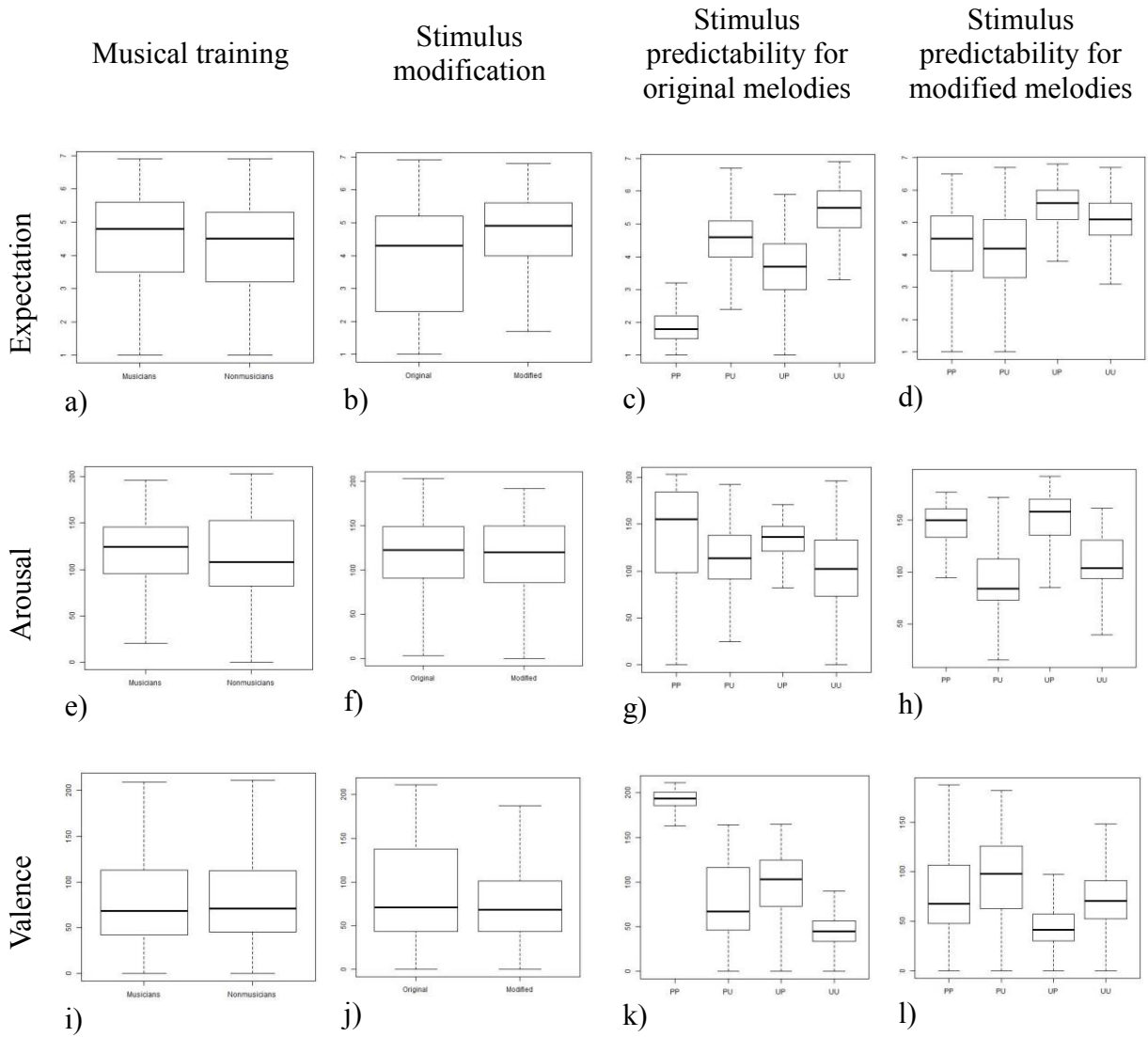

Figure 3. Box plots illustrating important mean comparisons between musicians and non-musicians (a, e, i), original and modified melodies (b, f, j), stimulus predictability categories for original (c, g, k) and modified (d, h, l) melodies for expectation (a, b, c, d), arousal (e, f, g, h) and valence (i, j, k, l) ratings.



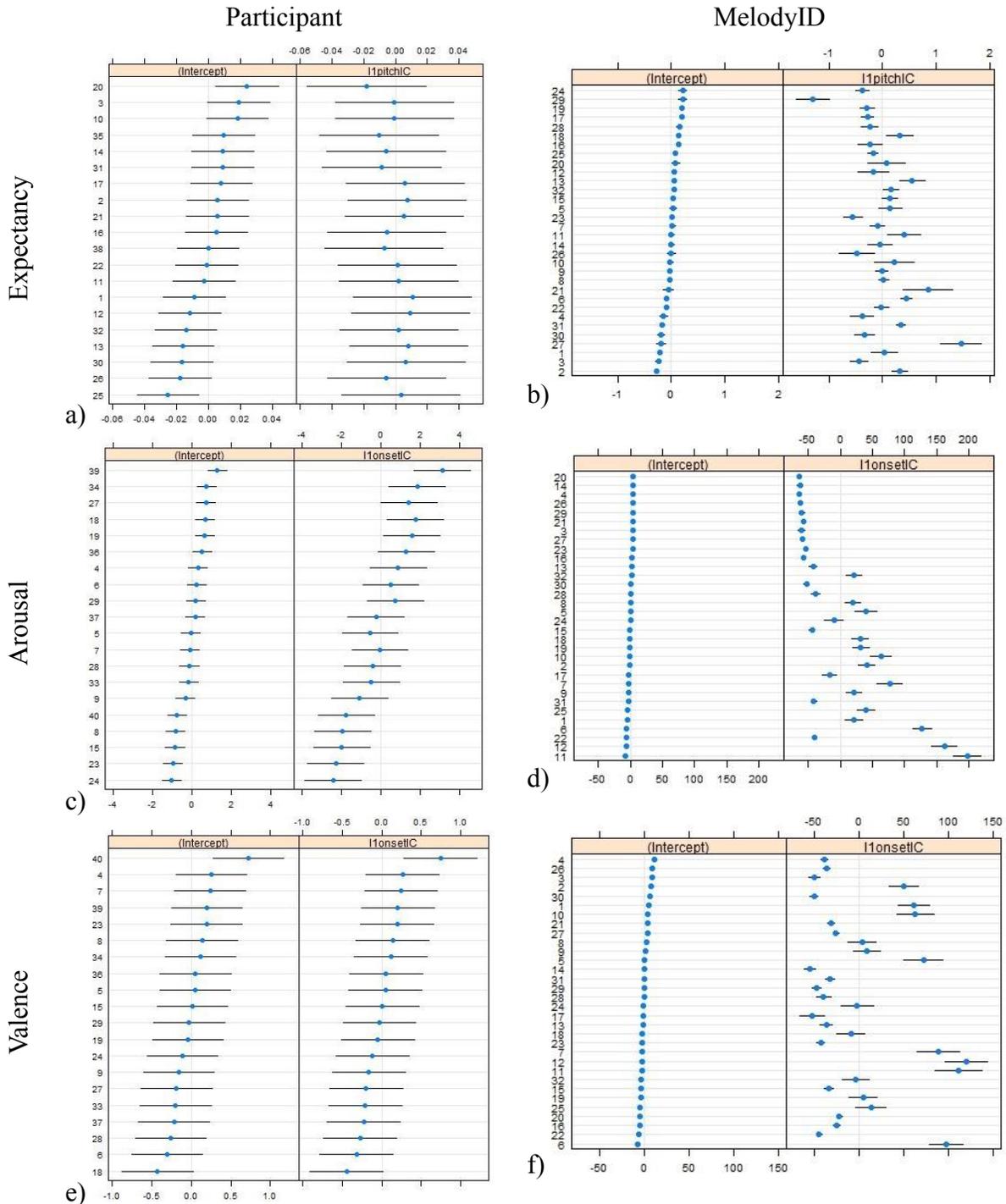

Figure 4. Intercept (left) and slope (right) values of random effects on Participant and MelodyID for expectancy, arousal and valence models. These show how each individual participant and melody was modelled and illustrate the variance among participants and melodies.



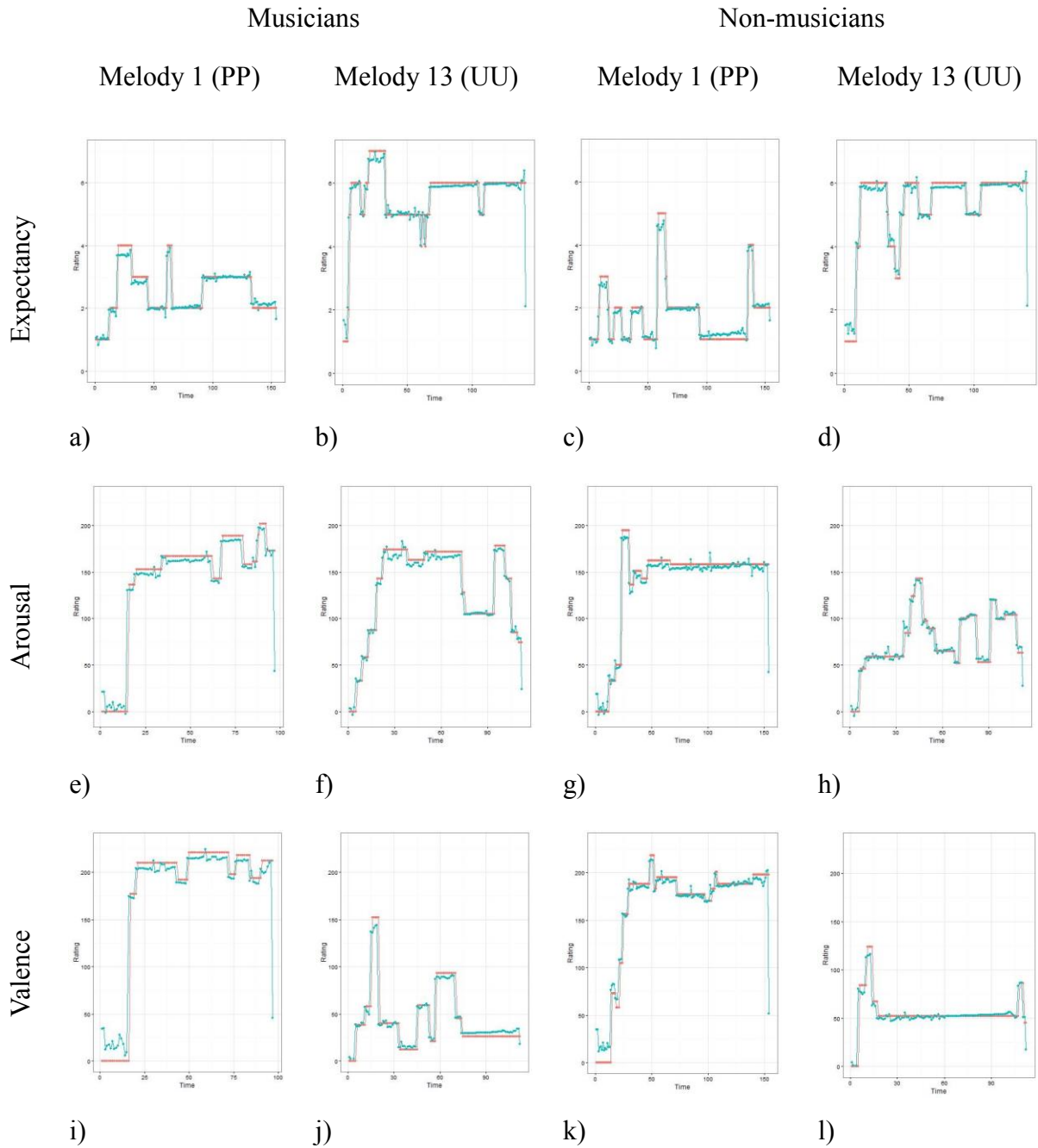

Figure 5. Expectancy (a, b, c, d), arousal (e, f, g, h) and valence (i, j, k, l) ratings for single randomly selected participants (6 musicians (a, b, e, f, i, j; participants 14, 35, 34, 18, 27, 7) and 6 non-musicians (c, d, g, h, k, l; participants 1, 10, 8, 33, 5, 37)) are plotted for Melodies 1 (a, c, e, g, i, k) and 13 (b, d, f, h, j, l), examples of PP and UU categories respectively. Ratings predicted by the model (teal) for those melodies for each of those participants only (single



extracts) are plotted alongside their actual ratings (pink).  Residuals were too small to illustrate on the same plot.



Appendix A

**Table A1**. Summary of 16 original melodies used in this experiment.

| File name | Dataset of origin | Number of events | Average pitch (60 = midC) | Average note duration (24 = quarter) | Mean pitch IC | Mean onset IC | Stimulus Predictability |
|-----------|-------------------|------------------|---------------------------|--------------------------------------|---------------|---------------|-------------------------|
| Kindr138 | 15 | 33 | 67.69 | 74.18 | 1.3624 | .8962 | PP |
| A162 | 18 | 21 | 70.23 | 27.42 | 1.4328 | .8955 | PP |
| Kindr151 | 15 | 51 | 66.05 | 22.82 | 1.5971 | .8114 | PP |
| Kindr162 | 15 | 19 | 68.36 | 26.52 | 1.5574 | .9254 | PP |
| Deut3480 | 10 | 19 | 72.89 | 36.94 | 2.4272 | 4.4488 | PU |
| Jugos052 | 4 | 54 | 66.22 | 6.77 | 2.2543 | 3.7433 | PU |
| I0511 | 23 | 53 | 66.83 | 11.67 | 2.0089 | 2.4660 | PU |
| Deut3284 | 10 | 67 | 69.71 | 6.52 | 2.0913 | 2.5380 | PU |
| I0533 | 23 | 39 | 67.76 | 11.79 | 5.6137 | 1.1401 | UP |
| A120 | 18 | 35 | 64.05 | 17.31 | 5.2750 | 1.3358 | UP |
| Oestr045 | 6 | 30 | 68.90 | 36.40 | 4.7200 | 1.1290 | UP |
| Oestr046 | 6 | 35 | 64.40 | 32.22 | 4.6734 | 1.1983 | UP |
| Deut3244 | 10 | 39 | 67.64 | 21.84 | 3.0216 | 4.7589 | UU |
| Deut3206 | 10 | 52 | 68.67 | 22.15 | 2.9122 | 4.5098 | UU |
| Deut3437 | 10 | 29 | 71.62 | 19.86 | 3.0114 | 4.3796 | UU |
| Deut3524 | 10 | 38 | 72.76 | 15.15 | 2.8472 | 4.3009 | UU |



Appendix B

**Table B1**. CSTSA modelling of expectancy ratings for all melodies; coefficients for fixed width 95% CI's and variance of random effects.

|  | Predictor | Coefficient | 95% CI | 95% CI |
|---|---|---|---|---|
|  | Intercept | .307 | .251 | .364 |
|  | Time | .0016 | .0015 | .0017 |
|  | Musicianship | .030 | .014 | .047 |
|  | Pitch | -.0022 | -.0026 | -.0019 |
|  | IOI | .0034 | .0031 | .0038 |
|  | L1ratings | .960 | .953 | .967 |
|  | L2ratings | -.065 | -.073 | -.058 |
|  | L4ratings | -.061 | -.069 | -.053 |
|  | L5ratings | .015 | .006 | .025 |
| **Fixed effects** | L6ratings | .035 | .023 | .037 |
|  | L15ratings | .015 | .012 | .018 |
|  | PitchIC | -.263 | -.309 | -.217 |
|  | L1pitchIC | .486 | .306 | .666 |
|  | L7pitchIC | .123 | .079 | .167 |
|  | L8pitchIC | -.059 | -.103 | -.016 |
|  | OnsetIC | -.731 | -.794 | -.667 |
|  | L1onsetIC | .845 | .769 | .920 |
|  | L2onsetIC | -.181 | -.240 | -.123 |
|  | L10onsetIC | -.084 | -.129 | -.039 |
|  | L12onsetIC | .138 | .092 | .183 |
|  | Predictor | Variance | - | - |
| **Random effects on individuals** | Intercept | .0002 |  |  |
|  | L1pitchIC | .0004 |  |  |
| **Random effects on melody ID** | Intercept | .019 |  |  |
|  | L1pitchIC | .245 |  |  |
| **Residual variance** |  | .421 |  |  |

**Table B2**. CSTSA modelling of arousal ratings for all melodies; coefficients for fixed width 95% CI's and variance of random effects.

|  | Predictor | Coefficient | 95% CI | 95% CI |
|---|---|---|---|---|
|  | (Intercept) | 1.98 | -.072 | 4.03 |
|  | Time | .067 | .064 | .070 |
|  | Predict2 | -3.42 | -5.77l | -1.06 |
|  | Predict3 | -.504 | -2.86 | 1.85 |
| **Fixed effects** | Predict4 | -4.53 | -6.88 | -2.17 |
|  | Pitch | -.045 | -.052 | -.038 |
|  | IOI | -.030 | -.037 | -.024 |
|  | L1ratings | .952 | .947 | .957 |
|  | L3ratings | .011 | .004 | .017 |
|  | L5ratings | -.059 | -.067 | -.051 |



| | Predictor | | | |
|---|---|---|---|---|
| | L6ratings | .032 | .025 | .039 |
| | L15ratings | .014 | .011 | .017 |
| | PitchIC | -16.6 | -17.7 | -15.4 |
| | L1pitchIC | 16.6 | 15.5 | 17.8 |
| | L6pitchIC | 2.46 | 1.31 | 3.62 |
| | L7pitchIC | 2.05 | .707 | 3.39 |
| | L8pitchIC | -2.14 | -3.37 | -.923 |
| | L10pitchIC | 1.86 | .633 | 3.08 |
| | L11pitchIC | -4.43 | -5.77 | -3.10 |
| | L12pitchIC | 4.91 | 3.57 | 6.25 |
| | L13pitchIC | -1.95 | -3.18 | -.725 |
| | L15pitchIC | 2.18 | 1.23 | 3.13 |
| | OnsetIC | -11.4 | -12.9 | -9.83 |
| | L1onsetIC | 72.4 | 48.2 | 96.6 |
| | L3onsetIC | 6.96 | 5.26 | 8.66 |
| | L4onsetIC | -8.38 | -9.98 | -6.77 |
| | L7onsetIC | 1.55 | .345 | 2.76 |
| | L10onsetIC | -6.81 | -8.12 | -5.49 |
| | L12onsetIC | 5.43 | 3.73 | 7.13 |
| | L13onsetIC | 4.47 | 2.55 | 6.39 |
| | L14onsetIC | -2.93 | -4.83 | -1.04 |
| | L15onsetIC | 3.09 | 1.59 | 4.58 |
| | **Predictor** | **Variance** | - | - |
| **Random effects on individuals** | Intercept | .474 | | |
| | L1onsetIC | 2.94 | | |
| **Random effects on melody ID** | Intercept | 13.5 | | |
| | L1onsetIC | 4815.2 | | |
| **Residual variance** | | 276.7 | | |

**Table B3**. CSTSA modelling of valence ratings for all melodies; coefficients for fixed width 95% CI's and variance of random effects.

| | Predictor | Coefficient | 95% CI | 95% CI |
|---|---|---|---|---|
| | (Intercept) | 5.38 | 3.56 | 7.20 |
| | Time | .037 | .033 | .042 |
| | Pitch | -.091 | -.102 | -.081 |
| | IOI | .167 | .153 | .181 |
| | L1ratings | .929 | .924 | .934 |
| | L3ratings | -.025 | -.033 | -.017 |
| **Fixed effects** | L4ratings | -.033 | -.042 | -.023 |
| | L5ratings | -.012 | -.022 | -.003 |
| | L6ratings | .017 | .008 | .027 |
| | L7ratings | .015 | .007 | .023 |
| | L9ratings | .008 | .003 | .014 |
| | L15ratings | .007 | .004 | .011 |
| | PitchIC | -9.19 | -10.6 | -7.72 |



| | | | |
|---|---|---|---|
| L1pitchIC | 11.2 | 9.74 | 12.6 |
| L5pitchIC | 2.62 | 1.45 | 3.79 |
| L8pitchIC | -3.26 | -4.72 | -1.79 |
| L9pitchIC | 3.29 | 1.74 | 4.83 |
| L11pitchIC | -1.68 | -3.22 | -.15 |
| L12pitchIC | 2.91 | 1.47 | 4.83 |
| L15pitchIC | 1.28 | .205 | 2.36 |
| OnsetIC | -20.0 | -22.2 | -17.9 |
| L1onsetIC | 48.5 | 29.7 | 67.3 |
| L3onsetIC | 4.05 | 1.92 | 6.18 |
| L4onsetIC | -4.02 | -5.90 | -2.13 |
| L10onsetIC | -5.35 | -6.65 | -4.05 |
| L13onsetIC | 3.59 | 2.32 | 4.86 |
| **Predictor** | **Variance** | - | - |
| **Random effects on individuals** | (Intercept) | .115 | |
| | L1onsetIC | .123 | |
| **Random effects on melody ID** | (Intercept) | 22.2 | |
| | L1onsetIC | 2878.7 | |
| **Residual variance** | | 439.9 | |



Appendix C

Perceptual salience is explained in a variety of ways in the current literature, and there is

currently no consensus on the correct way to describe, or measure it.  Dibben (1999) describes

salience in relation to pitch register, parallelism, and stability, Collins et al. (Collins, Laney,

Willis, & Garthwaite, 2011) and Huron (Huron, 2001) in terms of repetition, Prince et al. (Prince

et al., 2009) in terms of complexity, defined by number of different possible values (i.e. different

pitches or different rhythmic durations) and Lerdahl (1989), as a set of conditions combining

pitch register, timing, timbre, attack, note density, motivic content and grouping.  Outside of

music, salience is defined by predictability and uncertainty (Esber & Haselgrove, 2011), where

there are two possibilities: predictable content becomes more (i.e. a cue becomes salient if it

predicts a reward) or less (i.e. new information is more interesting) salient.  Where we interpret

larger CSTSA model coefficients to reflect more salient predictors, here we test the hypothesis

that melodies with high pitch predictability (expectancy) and low onset predictability (PU) have

larger pitch coefficients than onset coefficients, and vice versa.  To do so four sub-models of

each of the three CSTSA models optimised in the main experiment were created, one for each

category (PP, PU, UP, UU) in order to compare coefficients between models.  Details of these

models can be found in Tables C1-3.  A linear multiple regression model with stimulus

predictability, lag type (pitch, onset) and rating type (expectancy, arousal, valence) predicting the

coefficients of these CSTSA models revealed no significant effects, $F_{(3, 168)} = .50$, $p = .67$, $F_{(1, 170)} = 2.23$, $p = .13$, $F_{(2, 169)} = .51$, $p = .59$ respectively.  There were also no interactions

between category and lag type, $F_{(7, 164)} = .79$, $p = .59$.  While there was no statistically

significant effect, we observe that the sum of lags of onsetIC were consistently larger than the

sum of lags of pitchIC for all categories of stimulus predictability for arousal and valence



models, while the sum of lags of pitchIC were slightly larger than the sum of lags of onsetIC in the expectancy model.  In conclusion, our hypothesis was not supported here, where salience does not seem to be related to expectancy.  However, this study was not designed to investigate this question, which would be interesting to explore in a future study.



Table E1. Coefficients of sub-models for expectancy ratings.

|  | Coefficient | PP | PU | UP | UU |
|---|---|---|---|---|---|
| Fixed effects | Intercept | .393 | .399 | .235 | .204 |
|  | Time | .002 | .001 | .003 | .001 |
|  | Musicianship | .035 | .031 | .024 | .031 |
|  | Pitch | -.001 | .002 | -.002 | -.002 |
|  | IOI | .001 | .003 | .004 | .005 |
|  | L1ratings | .777 | 1.00 | 1.01 | 1.03 |
|  | L2ratings | .042 | -.100 | -.141 | -.110 |
|  | L4ratings | -.045 | -.075 | -.075 | -.035 |
|  | L5ratings | -.010 | -.031 | .031 | .021 |
|  | L6ratings | .052 | -.007 | .039 | .008 |
|  | L15ratings | .027 | .013 | .019 | .009 |
|  | PitchIC | -.192 | -.197 | -.142 | -.549 |
|  | L1pitchIC | .197 | .461 | .604 | .842 |
|  | L7pitchIC | .242 | .210 | -.021 | .008 |
|  | L8pitchIC | -.005 | -.169 | .030 | -.087 |
|  | OnsetIC | -.756 | -1.26 | -.316 | -.769 |
|  | L1onsetIC | 1.15 | 1.50 | .687 | .277 |
|  | L2onsetIC | -.258 | -.469 | -.231 | .078 |
|  | L10onsetIC | -.530 | -.169 | -.038 | -.153 |
|  | L12onsetIC | .188 | .034 | .058 | .235 |
| Random effects | Participant – Intercept | .002 | .0008 | .002 | 0.000 |
|  | Participant – l1pitchIC | .014 | .003 | .0009 | 3.001e-12 |
|  | MelodyID – Intercept | .105 | .005 | .010 | .006 |
|  | MelodyID – l1pitchIC | .078 | .174 | .337 | .178 |
| Residual variance |  | .455 | .397 | .536 | .319 |



Table E2. Coefficients of sub-models for arousal ratings

|  | Coefficients | PP | PU | UP | UU |
|---|---|---|---|---|---|
| Fixed effects | Intercept | -5.66 | .772 | -4.68 | -3.45 |
|  | Time | .255 | .040 | .246 | .058 |
|  | Pitch | -.027 | -.024 | -.104 | -.002 |
|  | IOI | -.030 | .002 | -.124 | -.001 |
|  | L1ratings | .870 | .965 | .894 | .954 |
|  | L3ratings | .027 | .010 | .012 | -.0006 |
|  | L5ratings | -.051 | -.083 | -.038 | -.028 |
|  | L6ratings | .033 | .047 | .044 | .019 |
|  | L15ratings | .053 | .013 | .028 | .016 |
|  | PitchIC | -18.3 | -12.7 | -11.6 | -20.5 |
|  | L1pitchIC | 19.8 | 8.85 | 14.5 | 24.1 |
|  | L6pitchIC | 9.35 | 1.42 | -1.65 | .329 |
|  | L7pitchIC | -2.64 | 3.32 | 8.12 | 2.05 |
|  | L8pitchIC | -.201 | -3.68 | -3.71 | -4.12 |
|  | L10pitchIC | 6.03 | .665 | 2.15 | .081 |
|  | L11pitchIC | -10.0 | -6.78 | 2.50 | -1.29 |
|  | L12pitchIC | 6.54 | 8.64 | -2.35 | 1.75 |
|  | L13pitchIC | .088 | -7.88 | 3.12 | -.454 |
|  | L15pitchIC | 3.69 | .046 | 2.14 | 5.71 |
|  | OnsetIC | -24.7 | -21.1 | 3.90 | -7.01 |
|  | L1onsetIC | 78.4 | 91.46 | 123.5 | 17.4 |
|  | L3onsetIC | 10.2 | 1.84 | 10.5 | 4.73 |
|  | L4onsetIC | -15.0 | -7.82 | -7.92 | -2.71 |
|  | L7onsetIC | -3.79 | 7.51 | .017 | -1.55 |
|  | L10onsetIC | -24.2 | -1.00 | -6.14 | -1.82 |
|  | L12onsetIC | 13.5 | -3.32 | 5.03 | 6.53 |
|  | L13onsetIC | 6.51 | 11.9 | 1.76 | .479 |
|  | L14onsetIC | -3.12 | -6.33 | -4.64 | -.819 |
|  | L15onsetIC | 1.30 | 1.28 | 5.12 | 2.40 |
| Random effects | Participant – Intercept | .449 | 1.17 | .120 | .724 |
|  | Participant – l1onsetIC | .516 | .086 | .241 | 1.10 |
|  | MelodyID – Intercept | 35.3 | 8.23 | 56.4 | 3.90 |
|  | MelodyID – l1onsetIC | 2443.5 | 3846.0 | 11190.0 | 447.2 |
| Residual variance |  | 368.1 | 225.3 | 323.7 | 186.2 |



Table E3. Coefficients of sub-models for valence ratings.

| | Coefficients | PP | PU | UP | UU |
|---|---|---|---|---|---|
| Fixed effects | Intercept | -2.42 | 9.68 | 1.35 | -1.88 |
| | Time | .169 | .005 | .174 | .051 |
| | Pitch | -.065 | -.114 | -.135 | -.0001 |
| | IOI | .176 | .141 | .235 | .143 |
| | L1ratings | .900 | .946 | .886 | .925 |
| | L3ratings | -.004 | -.018 | -.052 | -.009 |
| | L4ratings | -.024 | -.015 | -.054 | -.029 |
| | L5ratings | -.024 | -.041 | .023 | -.015 |
| | L6ratings | .021 | .028 | .0004 | .017 |
| | L7ratings | .002 | .00009 | .039 | .004 |
| | L9ratings | .017 | .003 | .026 | .006 |
| | L15ratings | .013 | .005 | .026 | .018 |
| | PitchIC | -8.42 | -6.56 | -7.56 | -14.7 |
| | L1pitchIC | 12.9 | 8.24 | 6.30 | 19.05 |
| | L5pitchIC | 2.98 | -.137 | 3.96 | 3.18 |
| | L8pitchIC | -1.07 | -3.18 | -3.32 | -8.26 |
| | L9pitchIC | 3.10 | 5.09 | 1.15 | 5.82 |
| | L11pitchIC | -.546 | -7.44 | 2.52 | -.533 |
| | L12pitchIC | 4.25 | 1.02 | 2.46 | 2.62 |
| | L15pitchIC | 1.71 | -.696 | 2.30 | 2.92 |
| | OnsetIC | -22.7 | -24.0 | -4.81 | -21.2 |
| | L1onsetIC | 49.1 | 80.7 | 81.4 | 6.13 |
| | L3onsetIC | 17.6 | 3.28 | .225 | 3.28 |
| | L4onsetIC | -13.1 | -.275 | -1.83 | -2.87 |
| | L10onsetIC | -14.3 | -4.66 | -6.26 | .210 |
| | L13onsetIC | 7.75 | 6.75 | -.153 | -.945 |
| Random effects | Participant − Intercept | .030 | .720 | .108 | .202 |
| | Participant − l1onsetIC | .147 | .001 | 1.37 | 1.06 |
| | MelodyID − Intercept | 46.5 | 9.41 | 49.8 | 4.26 |
| | MelodyID − l1onsetIC | 2972.0 | 4176.0 | 5024.3 | 36.0 |
| Residual variance | | 519.6 | 375.9 | 712.1 | 251.5 |